\documentclass[epjST]{svjour}
\usepackage{graphics,graphicx,cite}
\usepackage{amsfonts,verbatim,color}
\usepackage{amsmath,amssymb,slashed,wasysym}
\usepackage[T1]{fontenc}

\begin{document}

\newcommand{\tr}{\mathop{\mathrm{tr}}}
\newcommand{\bsigma}{\boldsymbol{\sigma}}
\newcommand{\bphi}{\boldsymbol{\phi}}
\newcommand{\re}{\mathop{\mathrm{Re}}}
\newcommand{\im}{\mathop{\mathrm{Im}}}
\renewcommand{\b}[1]{{\boldsymbol{#1}}}
\newcommand{\diag}{\mathrm{diag}}
\newcommand{\sign}{\mathrm{sign}}
\newcommand{\sgn}{\mathop{\mathrm{sgn}}}
\newcommand{\half}{\textstyle{\frac{1}{2}}}

\title{Quantum phase transitions in Dirac fermion systems}
\author{
Rufus Boyack\inst{1}\fnmsep\inst{2} \and
Hennadii Yerzhakov\inst{1} \and
Joseph Maciejko\inst{1}\fnmsep\inst{2}\fnmsep\thanks{\email{maciejko@ualberta.ca}}
}
\institute{
Department of Physics, University of Alberta, Edmonton, Alberta T6G 2E1, Canada \and
Theoretical Physics Institute, University of Alberta, Edmonton, Alberta T6G 2E1, Canada
}
\abstract{A key problem in the field of quantum criticality is to understand the nature of quantum phase transitions in systems of interacting itinerant fermions, motivated by experiments on a variety of strongly correlated materials. Much attention has been paid in recent years to two-dimensional (2D) materials in which itinerant fermions acquire a pseudo-relativistic Dirac dispersion, such as graphene, topological insulator surfaces, and certain spin liquids. This article reviews the phenomenology and theoretical description of quantum phase transitions in systems of 2D Dirac fermions.} 
\maketitle
\section{Introduction}
\label{sec:intro}

One of the foremost goals of modern condensed matter physics is to understand the equilibrium properties of many-particle systems at ultralow temperatures, where quantum effects dominate and classical notions become insufficient. The problem is most sharply formulated at zero temperature, where the notion of a thermodynamic phase of matter is replaced by that of a quantum phase of matter---an equivalence class of ground states of many-particle Hamiltonians with local interactions, and the notion of a thermal phase transition is supplanted by that of a quantum phase transition, driven by zero-point fluctuations. Systems of fermions, objects with no classical counterparts, are perhaps the most intriguing in this regard. Quantum phases of fermionic matter can be divided into two classes: gapped phases possess a nonzero gap to excitations above the (possibly degenerate) ground state that persists in the thermodynamic limit, while such a gap vanishes in a gapless phase. This dichotomy is best understood in the noninteracting limit, in which a gapped phase is a (possibly topological) band insulator, and a gapless phase is a metal or a semimetal. While noninteracting gapped phases are inherently stable against interactions, metals and semimetals are naturally susceptible to them, due to the gaplessness of their low-energy spectrum. When sufficiently strong, interactions can lead to a dramatic and sudden reorganization of a metallic or semimetallic ground state---a quantum phase transition. Experimentally relevant examples include the Mott metal-insulator transition~\cite{imada1998}, ferromagnetic or antiferromagnetic transitions in heavy fermion metals~\cite{si2010}, and possible quantum critical points (QCPs) in the phase diagram of high-temperature superconductors~\cite{sachdev2011}, all occurring in conventional metals with a nonrelativistic dispersion relation for electrons and the resulting extended Fermi surface.

This article reviews the physics of quantum phase transitions in a different class of itinerant Fermi systems: two-dimensional (2D) fermionic Dirac matter, where the dispersion relation of fermionic quasiparticles is linear at low energies and mimics that of massless Dirac fermions in (2+1)D relativistic quantum field theory. We begin by reviewing how and where massless Dirac fermions can arise in condensed matter systems (Sec.~\ref{sec:DiracCMP}). We then explain how quantum phase transitions can generally occur in those systems and how they may be described theoretically (Sec.~\ref{sec:QPT}). The remaining sections focus on specific examples of such transitions. We first discuss transitions from a Dirac semimetal to an insulator (Sec.~\ref{sec:SM-I}) or to a gapped superconductor (Sec.~\ref{sec:SUSY}). The ensuing two sections focus on more specialized topics. Section~\ref{sec:disorder} discusses possible effects of disorder at Dirac QCPs, and Sec.~\ref{sec:DSL} focuses on quantum phase transitions out of the Dirac spin liquid. We briefly conclude in Sec.~\ref{sec:conclusion}.

\section{Dirac fermions in condensed matter physics}
\label{sec:DiracCMP}

While all matter is believed to be described by a relativistic quantum field theory, the Standard Model of particle physics, at energies of interest in condensed matter physics ($\lesssim 1$~eV), matter is adequately described by a many-particle nonrelativistic Schr\"odinger equation for electrons. At such energies, the crystalline ionic lattice leads to a significant restructuring of the electron dispersion relation from its free-particle form, and physical properties at temperatures much less than the Fermi energy $E_F$ (we use natural units, $\hbar\!=\!k_B\!=\!1$) are largely controlled by the form of the dispersion relation near $E_F$. In one spatial dimension (1D), an arbitrary band dispersion relation $E(k)$, with $k$ the electron momentum, will generically cross $E_F$ linearly at $2N$ points in the first Brillouin zone $-\frac{\pi}{a}\!<k\!\leq\frac{\pi}{a}$ ($a$ is the lattice constant). Since $E(k)$ has the periodicity of the reciprocal lattice, this number of crossings is necessarily even, with an equal number of crossings with positive/negative group velocities. At low energies, the system is thus described by $N$ flavors of (1+1)D right-handed $\psi_{R,i}$ and left-handed $\psi_{L,i}$ Weyl fermions, $i\!=\!1,\ldots,N$, which can be combined into $N$ massless two-component Dirac fermions $\psi_i\!=\!(\psi_{R,i},\psi_{L,i})$.

While the appearance of massless Dirac fermions is generic in 1D, their equivalent in 2D---linear point crossings at the Fermi energy---can only occur as the result of a twofold band degeneracy. Consider the probability that two energy bands, given by the eigenvalues of an effective $2\times 2$ Bloch Hamiltonian $H(\b{k})$, become degenerate. Such a Hamiltonian can be generally written as $H(\b{k})\!=\!d_0(\b{k})+\b{d}(\b{k})\cdot\b\tau$, where $\b\tau\!=\!(\tau_x,\tau_y,\tau_z)$ is a vector of Pauli matrices and $d_0$, $\b{d}\!=\!(d_x,d_y,d_z)$ are real functions of $\b{k}$. The spectrum is $E_\pm(\b{k})\!=\!d_0\pm|\b{d}|$; $d_0$ preserves a twofold degeneracy of the bands for all $\b{k}$ and is inessential to our argument, while $\b{d}$ lifts this degeneracy unless all three of its components vanish. In 2D, degeneracy points are thus points of common intersection of three curves $d_\alpha(k_x,k_y)\!=\!0$, $\alpha\!=\!x,y,z$, in the $(k_x,k_y)$ plane~\cite{vonneumann1929}; such a problem has no solution in general. Put differently, by performing a suitable rotation of the Pauli matrices and a rotation/rescaling of momentum coordinates, near a putative twofold crossing at $\b{k}\!=\!\b{k}_0$ the Hamiltonian can be written as
\begin{align}\label{HDirac}
H(\b{k})\approx v_F(p_x\tau_x+p_y\tau_y)+m\tau_z,
\end{align}
to linear order in $\b{p}\!=\!\b{k}-\b{k}_0$, where $v_F$ is the Fermi velocity. This is the (2+1)D Dirac Hamiltonian, albeit with a mass $m$ that is generically nonzero. Thus massless Dirac fermions are not generic in 2D, but can appear if there is a discrete symmetry that forbids this mass term, as occurs in certain materials.

\begin{figure}
\centering
\includegraphics[width=0.7\columnwidth]{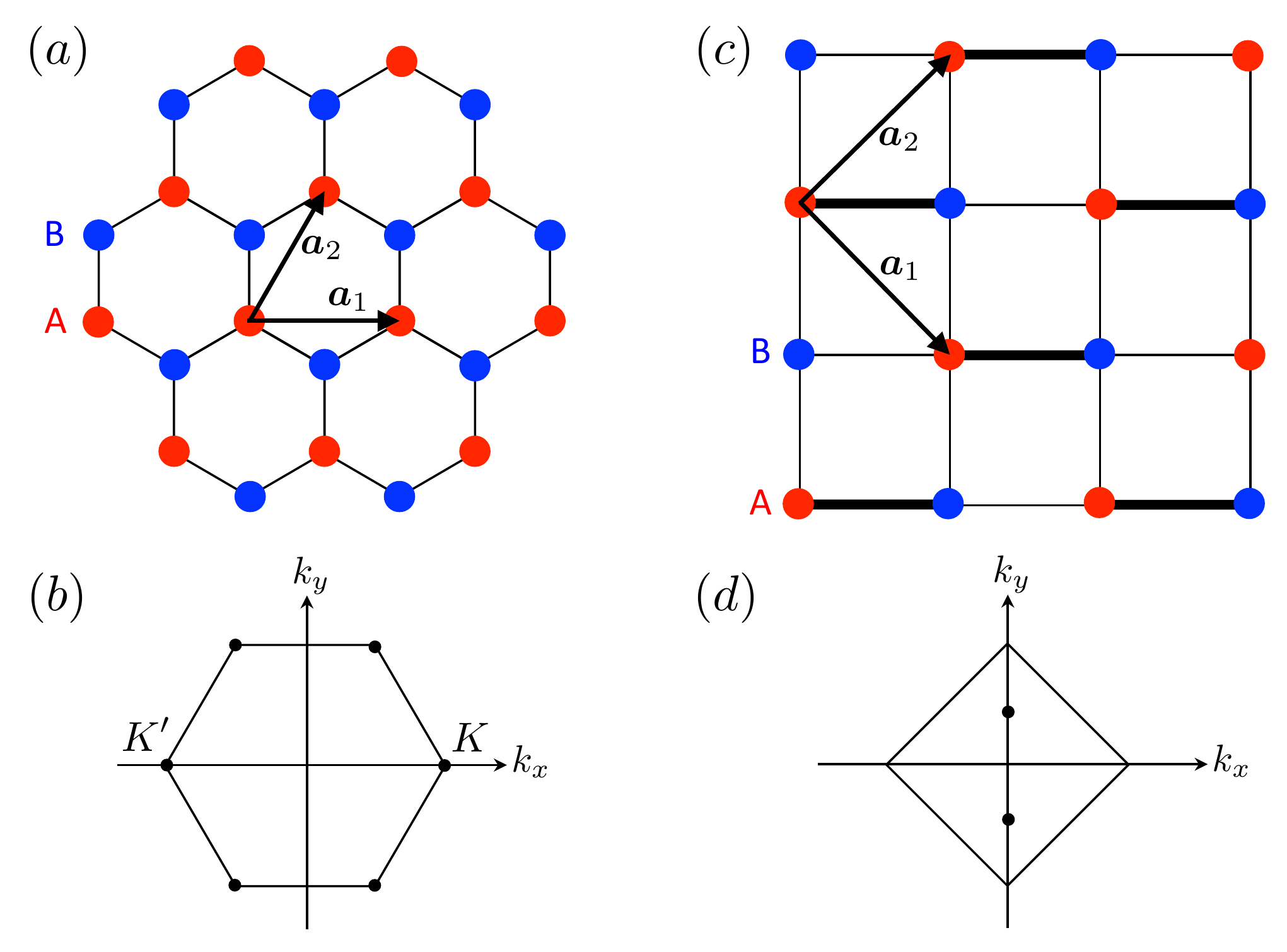}
\caption{(a) Honeycomb lattice of graphene and (b) its first Brillouin zone; (c) $\pi$-flux square lattice and (d) its first Brillouin zone.}
\label{fig:1}       
\end{figure}

The prototypical example of fermionic Dirac matter is electrons in graphene~\cite{wallace1947,novoselov2005,zhang2005}, an atomic monolayer of carbon atoms forming a 2D honeycomb lattice [Fig.~\ref{fig:1}(a)]. The bands crossing the Fermi energy are obtained by considering a model of electrons with nearest-neighbor hopping $t$ on this lattice, which corresponds to a triangular Bravais lattice with a unit cell comprising two (A and B) sites. The Bravais lattice vectors can be chosen as $\b{a}_1\!=\!(1,0)$ and $\b{a}_2\!=\!(\frac{1}{2},\frac{\sqrt{3}}{2})$ in Cartesian coordinates, setting the shortest A-A distance to unity. In the sublattice (AB) basis, the Hamiltonian reads
\begin{align}\label{TBHamiltonian}
H(\b{k})=-t\left(\begin{array}{cc}
0 & f(\b{k}) \\
f^*(\b{k}) & 0
\end{array}\right),
\end{align}
where $f(\b{k})\!=\!1\!+\!e^{i\b{k}\cdot(\b{a}_1-\b{a}_2)}\!+\!e^{-i\b{k}\cdot\b{a}_2}$. This Hamiltonian has twofold-degenerate linear crossings at the six corners of the Brillouin zone: $(\pm\frac{4\pi}{3},0)$, $(\pm\frac{2\pi}{3},\frac{2\pi}{\sqrt{3}})$, and $(\pm\frac{2\pi}{3},-\frac{2\pi}{\sqrt{3}})$ [black dots in Fig.~\ref{fig:1}(b)]. Only two such Dirac points or ``valleys'', e.g., $(\pm\frac{4\pi}{3},0)$, are inequivalent, and denoted by $K$ and $K'$, respectively. Expanding near those Dirac points as previously, and performing suitable basis rotations, we obtain $N\!=\!4$ copies (2 valleys $\times$ 2 spins) of the Dirac Hamiltonian (\ref{HDirac}) with $v_F\!=\!\sqrt{3}t/2$ and $m\!=\!0$. The vanishing of the Dirac mass is a consequence of inversion symmetry on the honeycomb lattice combined with spinless time-reversal symmetry ($\mathcal{T}$). Inversion about a hexagon center reverses the direction of momentum and interchanges the A and B sublattices, an operation effected by the $\tau_x$ Pauli matrix. $\mathcal{T}$ is an antiunitary symmetry that also flips the momentum, but additionally complex-conjugates $c$-numbers (spin is a spectator in this argument). Combining those operations, the Hamiltonian transforms as $H(\b{k})\!\rightarrow\!\tau_xH^*(\b{k})\tau_x$, which is a symmetry of (\ref{HDirac}) only if $m\!=\!0$.

These results only hold in the absence of spin-orbit coupling, which in fact opens a minuscule gap $\sim 10^{-3}$~meV at the Dirac points in graphene~\cite{kane2005,min2006,yao2007}. In the presence of spin-orbit coupling, protected Dirac points can nonetheless occur in 2D crystals with non-symmorphic space groups~\cite{young2015} or on the surfaces of 3D topological insulators~\cite{hasan2010,qi2011}. Dirac crossings in the latter case occur at $\mathcal{T}$-invariant momenta in the surface Brillouin zone, but are now a consequence of Kramers' degeneracy. The two components in the Hamiltonian (\ref{HDirac}) correspond in this case to spin up and spin down, and $\mathcal{T}$ acts as $H(\b{k})\!\rightarrow\!\tau_yH^*(-\b{k})\tau_y$, which again forbids a $\tau_z$ mass term. In contrast to strict 2D $\mathcal{T}$-invariant systems such as graphene, the surface of a 3D topological insulator can support an odd number of Dirac fermions, e.g., $N\!=\!1$ in Bi$_2$Se$_3$~\cite{xia2009} and Bi$_2$Te$_3$~\cite{chen2009}, $N\!=\!3$ in Zr$_2$Te$_2$P~\cite{ji2016} and LaBi~\cite{nayak2017}, and $N\!=\!5$ in Bi$_{1-x}$Sb$_x$~\cite{hsieh2008}.

Besides appearing in the bandstructure of noninteracting electrons, Dirac fermions can also appear as low-energy excitations in interacting systems, such as superconductors or spin liquids. Unconventional superconductors with $d_{x^2-y^2}$ spin-singlet pairing, like the cuprate high-temperature superconductors~\cite{miyake1986,scalapino1986}, support $N\!=\!4$ flavors of Bogoliubov quasiparticle excitations at low energies, each described by a two-component Nambu spinor and an anisotropic massless Dirac Hamiltonian. In the Dirac or algebraic spin liquid~\cite{affleck1988,marston1989,kim1999,rantner2001,rantner2002,hermele2004,hermele2005,hermele2007}---which may describe the ground state of the spin-1/2 Heisenberg model on the kagome lattice~\cite{hastings2000,ran2007,hermele2008,iqbal2011,iqbal2013,iqbal2014,he2017,zhu2018}, applicable to ZnCu$_3$(OH)$_6$Cl$_2$ (herbertsmithite~\cite{han2012,fu2015})---conventional spin-1 bosonic excitations, such as a single spin flip, ``fractionalize'' into spin-1/2 Dirac fermions coupled to a fluctuating $U(1)$ gauge field. A simple theoretical model for this state~\cite{affleck1988,marston1989} is given by fermions with nearest-neighbor hopping $t$ on the $\pi$-flux square lattice [Fig.~\ref{fig:1}(c)], and additionally coupled to a dynamical lattice $U(1)$ gauge field living on the links of the lattice. A background magnetic flux of $\pi$ through each square plaquette, obtained for instance by flipping the sign of the hopping amplitude on each thick bond in Fig.~\ref{fig:1}(c), produces two Dirac crossings in the bandstructure. For the choice $\b{a}_1\!=\!\hat{\b{x}}-\hat{\b{y}}$, $\b{a}_2\!=\!\hat{\b{x}}+\hat{\b{y}}$ of lattice vectors in Fig.~\ref{fig:1}(c), setting the bond length to unity, Brillouin zone corners occur at $(\pm \pi,0),(0,\pm\pi)$ and the fermion Hamiltonian is given again by Eq.~(\ref{TBHamiltonian}) but with $f(\b{k})\!=\!1\!-\!e^{i(k_x-k_y)}\!+\!e^{-i(k_x+k_y)}\!+\!e^{-2ik_y}$. The Dirac points are found at wavevectors $\pm\b{Q}\!=\!\pm(0,\frac{\pi}{2})$, shown as black dots in Fig.~\ref{fig:1}(d). Taking fluctuations of the lattice gauge field into account, at low energies the model reduces to (2+1)D quantum electrodynamics (QED$_3$) with $N\!=\!4$ flavors of massless two-component Dirac fermions.

\section{Quantum phase transitions}
\label{sec:QPT}

A quantum phase transition is a zero-temperature phase transition tuned by a non-thermal parameter, such as pressure, doping, or magnetic field, and accompanied by a singularity in the ground-state energy density~\cite{QPT}. To understand how a quantum phase transition can occur in any of the condensed matter systems just discussed, consider its description by means of a low-energy effective Lagrangian,
\begin{align}\label{LGN}
\mathcal{L}_\text{GN}=\overline{\Psi}\slashed{\partial}\Psi-g(\overline{\Psi}\b{\mathcal{M}}\Psi)^2,
\end{align}
where $\Psi\!=\!(\psi_1,\ldots,\psi_{N_f})$ is a generalized spinor containing all $N_f\!=\!N/2$ flavors of four-component Dirac fermions~\cite{NoteDirac}, $\overline{\Psi}\!=\!\Psi^\dag\gamma_0$ is its Dirac conjugate, and we have set the Fermi velocity to unity. Lagrangians of this type were first considered by Nambu and Jona-Lasinio (NJL)~\cite{NJL1961} and Gross and Neveu (GN)~\cite{gross1974} in high-energy physics. The first term describes the free Dirac dispersion, where $\slashed{\partial}\!=\!\gamma_\mu\partial_\mu$, $\mu\!=\!0,1,2$ denote Euclidean spacetime coordinates, and $\gamma_\mu$ is a choice of Dirac gamma matrices appropriate to the problem at hand. The second term describes two-body interactions among the Dirac fermions, with strength $g$. We assume interactions are dominant in a particular channel, represented by $n$ matrices $\mathcal{M}_A$, $A\!=\!1,\ldots,n$, which is generically true near a particular phase transition; the Lagrangian further has a global $O(n)$ symmetry. When $g\!=\!0$, one obtains an $O(n)$-symmetric phase with gapless free Dirac fermions. For very large $g\!>\!0$, the $O(n)$ order parameter $\overline{\Psi}\b{\mathcal{M}}\Psi$ acquires a ground-state expectation value, which breaks the $O(n)$ symmetry spontaneously. We also assume the structure of $\b{\mathcal{M}}$ is such that it corresponds to a mass term for the Dirac fermions, opening a gap in their spectrum. While not the only possibility, it is often the most energetically favorable.
 
To see that this change must happen abruptly at some critical value $g\!=\!g_c$ of the interaction strength, we can study Eq.~(\ref{LGN}) in mean-field theory. We decouple the two-body interaction using an $O(n)$ Hubbard-Stratonovich field $\boldsymbol{\phi}$, whose equation of motion identifies it with the order parameter above. We then integrate out the fermions and perform a saddle-point approximation, assuming a constant $\boldsymbol{\phi}$. Assuming a symmetry-breaking solution $\boldsymbol{\phi}\neq 0$, we obtain a zero-temperature gap equation given schematically by
\begin{align}\label{GapEq}
\frac{1}{g}\propto\int^\Lambda\frac{d^2\b{p}}{(2\pi)^2}\int\frac{d\omega}{2\pi}\frac{1}{\omega^2+\b{p}^2+\boldsymbol{\phi}^2},
\end{align}
where $\Lambda\!\sim\!a^{-1}$ is a large momentum cutoff, with $a$ the lattice constant. Equation~(\ref{GapEq}) has a solution only for $g\!>\!g_c$, where $\frac{1}{g_c}\!\propto\!\int^\Lambda\frac{d^2\b{p}}{(2\pi)^2}\int\frac{d\omega}{2\pi}\frac{1}{\omega^2+\b{p}^2}\!\propto\!\Lambda$; thus the transition occurs at a finite interaction strength $g\!=\!g_c$.

In the vicinity of a (continuous) quantum phase transition such as this, physical observables obey scaling laws with universal critical exponents. The correlation length $\xi$ and the characteristic energy scale $\Delta$ (or alternatively, the correlation time $\xi_\tau\!\sim\!\hbar/\Delta$) depend on the distance to the critical point as
\begin{align}
\xi\sim|g-g_c|^{-\nu},\hspace{5mm}\Delta\sim\xi^{-z}\sim|g-g_c|^{z\nu},
\end{align}
where $\nu$ is the correlation length exponent and $z$ is the dynamic critical exponent. With relativistic Dirac fermions, $z\!=\!1$ generally since space and time scale in the same way, as can be seen from the like powers of $\omega$ and $\b{p}$ in the gap equation. (However, in Sec.~\ref{sec:disorder} we will discuss examples of Dirac QCPs with $z\!\neq\!1$.) In the ordered phase, $\Delta$ can be identified with the Dirac mass gap $\!\propto\!|\b{\phi}|$. Expanding (\ref{GapEq}) near $g\!=\!g_c$ reveals that $|\b{\phi}|\!\propto\!(g-g_c)$ for $g\!>\!g_c$, thus $\nu\!=\!1$ in this mean-field theory, which is in fact exact in the $N_f\!\rightarrow\!\infty$ limit~\cite{moshe2003}. Fluctuation corrections to this result can be computed systematically for GN models (\ref{LGN}) in powers of $1/N_f$, an approach known as the large-$N_f$ expansion~\cite{vasilev1981}. Alternatively, one can work with $N_f$ finite but formulate the theory in $d\!=\!4\!-\!\epsilon$ spacetime dimensions and compute critical exponents in the $\epsilon$ expansion~\cite{wilson1972}. The appropriate Lagrangian in the latter case is the Gross-Neveu-Yukawa (GNY) Lagrangian~\cite{zinn-justin1991},
\begin{align}\label{LGNY}
\mathcal{L}_\text{GNY}=\overline{\Psi}\slashed{\partial}\Psi+(\partial_\mu\b{\phi})^2+m^2\b{\phi}^2+\lambda^2(\b{\phi}^2)^2+h\b{\phi}\cdot\overline{\Psi}\b{\mathcal{M}}\Psi,
\end{align}
where all terms marginal or relevant in four dimensions and consistent with symmetries are included. The tuning parameter for the transition in this case is $m^2$, with $m^2>0$ ($m^2<0$) corresponding to the disordered (ordered) phase, and $\nu\!=\!\frac{1}{2}+\mathcal{O}(\epsilon)$. For theories that can be continued from three to four dimensions in a Lorentz-invariant manner, all known critical exponents match order by order in the large-$N_f$ and $\epsilon$ expansions, establishing the equivalence of the two formulations. As for metallic quantum criticality in 2+1 dimensions~\cite{abanov2004}, integrating out the fermions directly from (\ref{LGNY}) following the Hertz-Millis approach~\cite{hertz1976,millis1993} would yield an intractable effective action for $\b{\phi}$ with an infinite number of nonlocal vertices~\cite{strack2010}. By contrast, in the $\epsilon$ expansion fermions and bosons are treated on equal footing, and the theory (\ref{LGNY}) is perfectly local.

QCPs of Dirac fermions are also examples of non-Fermi liquids~\cite{QPT}. At criticality, the Dirac field interacts strongly with soft fluctuations of the bosonic order parameter $\b{\phi}$, and generally develops a nonzero anomalous dimension $\eta_\psi$. Additionally, $\b{\phi}$ itself acquires an anomalous dimension $\eta_\phi$ via self-interactions---as in the theory of classical critical phenomena---but also through interactions with the Dirac field. At criticality, the fermion $A(\b{p},\omega)$ and order parameter $\chi(\b{p},\omega)$ spectral functions acquire power-law divergences controlled by those anomalous dimensions,
\begin{align}
A(\b{p},\omega)\sim\frac{(\omega-\b{\alpha}\cdot\b{p})\Theta(\omega^2-\b{p}^2)\sgn\omega}{|\omega^2-\b{p}^2|^{1-\eta_\psi/2}},\hspace{5mm}
\chi(\b{p},\omega)\sim\frac{\Theta(\omega^2-\b{p}^2)\sgn\omega}{|\omega^2-\b{p}^2|^{1-\eta_\phi/2}},
\end{align}
where $\alpha_i\!=\!i\gamma_0\gamma_i$, $i\!=\!1,2$ and $\Theta(x)$ is the Heaviside step function. In the large-$N_f$ limit in fixed $d\!=\!3$ spacetime dimensions, one generally obtains $\eta_\psi\!=\!\mathcal{O}(1/N_f)$ and $\eta_\phi\!=\!1+\mathcal{O}(1/N_f)$~\cite{moshe2003}, while $\eta_\psi$ and $\eta_\phi$ are typically both $\mathcal{O}(\epsilon)$ in the $\epsilon$ expansion~\cite{zinn-justin1991}.

\section{Dirac semimetal-insulator transitions}
\label{sec:SM-I}

Given a condensed matter system with $N_f$ massless four-component Dirac fermions~\cite{NoteDirac} subject to sufficiently strong two-body interactions, a variety of long-range charge, spin, and superconducting orders are in principle possible, corresponding to the number of independent order parameters $\overline{\Psi}\b{\mathcal{M}}\Psi$ one can write down~\cite{herbut2009b,ryu2009} (for superconducting orders, the fields $\psi_i$ contained in $\Psi$ are Nambu spinors). Assuming the unbroken phase is a weakly interacting Dirac semimetal like graphene, we first discuss interaction-induced particle-hole (charge/spin) orders, where the broken phase is an insulator with single-particle gap $\Delta$. Superconducting orders are discussed in Sec.~\ref{sec:SUSY}. Symmetry-breaking orders that do not gap out the Dirac quasiparticles---like nematic order~\cite{xu2017,lundgren2017}---are also possible, as well as Lifshitz transitions caused by interaction-induced band renormalizations~\cite{dora2013}; neither will be discussed here.

\subsection{Charge-density-wave order and N\'eel antiferromagnetism}
\label{sec:CDW_AF}

In Sec.~\ref{sec:DiracCMP} we saw that Dirac crossings on the honeycomb lattice are protected by a combination of spinless $\mathcal{T}$ and inversion symmetries, which acts by complex conjugation together with A-B sublattice exchange. If this $\mathbb{Z}_2$ symmetry is broken spontaneously due to interactions, a mass gap can be dynamically generated. For repulsively interacting spinless fermions at half filling, a natural form of order is $\b{q}\!=\!0$ (staggered) charge-density-wave (CDW) order, with order parameter
\begin{align}\label{OCDW}
\mathcal{O}_\text{CDW}=\frac{1}{\mathcal{N}}\sum_\b{R}\left(c_{\b{R}A}^\dag c_{\b{R}A}-c_{\b{R}B}^\dag c_{\b{R}B}\right),
\end{align}
where $\b{R}\!=\!n_1\b{a}_1+n_2\b{a}_2$, $n_1,n_2\in\mathbb{Z}$ labels the $\mathcal{N}$ sites of the underlying triangular Bravais lattice, and $c_{\b{R}A(B)}^\dag$/$c_{\b{R}A(B)}$ creates/annihilates a fermion on sublattice A (B). In a phase with CDW long-range order, $\mathcal{T}$ is preserved, but sublattice symmetry is spontaneously broken. Since particle number is conserved, the resulting phase is an insulator, with massive Dirac fermions~\cite{semenoff1984}. This scenario is realized in models of spinless fermions with density-density interactions on the honeycomb lattice,
\begin{align}\label{HtV}
H_{t\text{-}V}=-t\sum_{\langle ij\rangle}(c_i^\dag c_j+\mathrm{h.c.})+V_1\sum_{\langle ij\rangle}\left(n_i-\half\right)\left(n_j-\half\right)+\ldots,
\end{align}
where $n_i\!=\!c_i^\dag c_i$ is the density operator on site $i$, $V_1\!>\!0$ is the nearest-neighbor repulsion, and $\ldots$ denotes possible longer-range interactions. In mean-field theory, the pure-$V_1$ model exhibits a quantum phase transition from a Dirac semimetal to a CDW insulator at a critical value of $V_1/t$~\cite{herbut2006,raghu2008,weeks2010,grushin2013}, a result confirmed by exact diagonalization (ED) studies on small clusters~\cite{garcia-martinez2013,capponi2015}; infinite density-matrix renormalization group (iDMRG) calculations~\cite{motruk2015}; the functional renormalization group (fRG) method~\cite{scherer2015}; and sign-problem-free quantum Monte Carlo (QMC) calculations, using the fermion bag~\cite{huffman2014,huffman2020}, continuous-time~\cite{wang2014,wang2015}, or Majorana-based~\cite{li2015a,li2015} algorithms. Similar results~\cite{wang2014,li2015} are found for the $\pi$-flux square lattice [Fig.~\ref{fig:1}(c)]. The appropriate critical theory (\ref{LGN},\ref{LGNY}) is the $N_f\!=\!1$ chiral Ising GN(Y) model~\cite{herbut2006,gross1974,zinn-justin1991}: a single real scalar field $\phi$, corresponding to the CDW order parameter, is Yukawa-coupled to the usual Dirac mass $\overline{\Psi}\Psi$. A transition in the same universality class is also found using a designer Hamiltonian that emulates the GNY Lagrangian itself, where noninteracting fermions directly couple to a dynamical real bosonic field on the lattice~\cite{liu2020}. Critical exponents for the chiral Ising universality class have been computed in both the large-$N_f$~\cite{zinn-justin1991,gracey1991,gracey1994} and $\epsilon$~\cite{zinn-justin1991,rosenstein1993,fei2016,mihaila2017,zerf2017,ihrig2018} expansions, fRG~\cite{janssen2014}, and the conformal bootstrap~\cite{iliesiu2016,iliesiu2018}. Besides critical exponents, other critical properties recently computed include the universal optical conductivity in the collisionless ($\omega\!\gg\!T$) regime~\cite{roy2018}, the universal low-energy spectrum~\cite{schuler2019}, and thermalization and quantum-chaotic properties~\cite{jian2018,jian2019}.

For spin-1/2 fermions, appropriate to electronic systems like graphene, a semimetal-CDW quantum phase transition can again be triggered by sufficiently strong nearest-neighbor~\cite{herbut2006,raghu2008,classen2014} or long-range~\cite{khveshchenko2001,gorbar2002,khveshchenko2006,khveshchenko2009,gamayun2010} electron-electron interactions, and also by sufficiently strong electron-phonon interactions, as shown by studies of the Holstein model on the honeycomb~\cite{chen2019,zhang2019} and $\pi$-flux~\cite{zhang2020} lattices. As charge order is diagonal in spin indices, this transition is expected to be again in the chiral Ising universality class, but with $N_f\!=\!2$. However, for spinful systems long-range order in the spin channel is also possible. On the bipartite lattices of Fig.~\ref{fig:1}, a natural candidate is N\'eel antiferromagnetic spin-density-wave (SDW) order, with order parameter
\begin{align}\label{OSDW}
\mathcal{O}_\text{SDW}=\frac{1}{\mathcal{N}}\sum_\b{R}\left(c_{\b{R}A}^\dag\bsigma c_{\b{R}A}-c_{\b{R}B}^\dag\bsigma c_{\b{R}B}\right),
\end{align}
where $\bsigma$ is a vector of spin Pauli matrices, and the fermion creation/annihilation operators now carry an implicit spin-1/2 index. For such fermions, on-site interactions $U$ are now possible, and the spinful analog to Eq.~(\ref{HtV}) is the half-filled Hubbard model,
\begin{align}\label{Hubbard}
H_\text{Hubbard}=-t\sum_{\langle ij\rangle \sigma}(c_{i\sigma}^\dag c_{j\sigma}+\mathrm{h.c.})+U\sum_i n_{i\uparrow}n_{i\downarrow}+\ldots,
\end{align}
where $n_{i\sigma}\!=\!c_{i\sigma}^\dag c_{i\sigma}$, $U\!>\!0$, and $\ldots$ denotes possible extended interactions. The model with purely on-site interactions has been studied extensively on the honeycomb and $\pi$-flux square lattices in mean-field theory~\cite{raghu2008}, sign-problem-free QMC~\cite{sorella1992,otsuka2002,sorella2012,assaad2013,otsuka2014,parisentoldin2015,otsuka2016,guo2018b,tang2018}, dynamical mean-field theory (DMFT)~\cite{raczkowski2020}, and fRG~\cite{honerkamp2008}. Additional nearest-neighbor~\cite{buividovich2018} or longer-range~\cite{ulybyshev2013} interactions have also been considered~\cite{GrapheneLongRange}. A continuous quantum phase transition from a Dirac semimetal to an insulating SDW state is found at a critical value of $U/t$. The critical theory is the chiral Heisenberg GN(Y) model---Eqs. (\ref{LGN},\ref{LGNY}) with $\b{\mathcal{M}}\!=\!\bsigma$ and $\b{\phi}$ a real $O(3)$ vector---with $N_f\!=\!1$ flavor of $SU(2)$ doublets of four-component Dirac fermions, i.e., eight complex fermionic degrees of freedom in total~\cite{herbut2006,herbut2009,roy2011}. (A model with a single $SU(2)$ doublet of {\it two}-component Dirac fermions is also allowed on a 2D lattice~\cite{NoteDirac,buttner2011}, and has been studied in Ref.~\cite{lang2019}.) Critical exponents for this model have been computed in fRG~\cite{janssen2014} and in the large-$N_f$~\cite{gracey2018} and $\epsilon$~\cite{rosenstein1993,zerf2017} expansions. Chiral Ising/Heisenberg {\it tricritical} points separating critical (chiral Ising/Heisenberg) and first-order semimetal-insulator phase boundaries have been shown to exist and argued to belong to distinct universality classes~\cite{yin2018}.

\subsection{Kekul\'e valence-bond-solid order and Haldane mass generation}
\label{sec:Kekule}

On the lattices of Fig.~\ref{fig:1}, the CDW and SDW orders discussed in Sec.~\ref{sec:CDW_AF} are $\b{q}\!=\!0$ orders which do not enlarge the unit cell. Finite-$\b{q}$ orders, which break symmetry under translations on the underlying Bravais lattice, are also possible. From a weak-coupling perspective, the most natural order of this type has $\b{q}$ nesting two inequivalent Dirac cones in the Brillouin zone, as in Kekul\'e valence-bond-solid (VBS) order on the honeycomb lattice~\cite{chamon2000,hou2007}. As in the Kekul\'e structure of benzene, a spontaneous modulation of the hopping strength alternates in magnitude from one bond to the next as one goes round an elementary hexagonal plaquette. Extended periodically to the lattice, this pattern triples the unit cell and breaks the discrete $C_3$ rotation symmetry of the honeycomb lattice. By contrast with $\b{q}\!=\!0$ orders, finite-$\b{q}$ nesting gaps out the Dirac cones pairwise, but likewise results in an insulator. Sign-problem-free QMC studies have shown that models with $SU(N_f)$ fermions on the honeycomb lattice with nearest-neighbor interactions~\cite{lang2013,li2017,li2019,li2020} or on-site Hubbard interactions~\cite{zhou2016} can exhibit a continuous quantum phase transition from a semimetal with $N_f\geq 1$ flavors of four-component Dirac fermions to an insulating Kekul\'e VBS phase, tuned by the interaction strength. This transition was also found in QMC studies of extended Hubbard models with cluster charge interactions~\cite{xu2018,daliao2019}, with possible applications to Mott physics in twisted bilayer graphene~\cite{po2018}, and an fRG study of an extended Hubbard model with electron-phonon interactions~\cite{classen2014}.

The critical theory of the Kekul\'e VBS transition for $SU(N_f)$ fermions is the chiral XY GN(Y) model with $N_f$ flavors of four-component Dirac fermions. In the language of Sec.~\ref{sec:QPT}, the order parameter is a real $O(2)$ vector field $\b{\phi}\!=\!(\phi_1,\phi_2)$, or equivalently a complex scalar field $\phi\!=\!\phi_1\!+\!i\phi_2$. By a suitable choice of Dirac matrices, the $O(2)$ doublet of matrices appearing in the corresponding fermion bilinear in Eqs.~(\ref{LGN},\ref{LGNY}) can be chosen as either $\b{\mathcal{M}}\!=\!(i\Gamma_3,i\Gamma_5)$~\cite{scherer2016} or $\b{\mathcal{M}}\!=\!(1,i\gamma_5)$~\cite{ryu2009}, where the two choices are related by $\gamma_\mu\!=\!i\Gamma_\mu\Gamma_3$, $\mu\!=\!0,1,2$, and $\gamma_5\!=\!-i\Gamma_3\Gamma_5$. The five matrices $\Gamma_\mu$, $\Gamma_3$, and $\Gamma_5\!=\!\Gamma_0\Gamma_1\Gamma_2\Gamma_3$ obey the $SO(5)$ Clifford algebra $\{\Gamma_a,\Gamma_b\}\!=\!2\delta_{ab}$. In the latter representation, Eq.~(\ref{LGN}) corresponds to the (2+1)D NJL model~\cite{NJL1961}, with the $O(2)$ symmetry of the former representation mapping onto a $U(1)$ chiral symmetry. Besides QMC, critical exponents for the chiral XY GN(Y) model have been computed in fRG~\cite{classen2017} and in the large-$N_f$~\cite{gracey1993,gracey1994b} and $\epsilon$~\cite{rosenstein1993,scherer2016,zerf2017,jian2017,roy2019} expansions.

In reality, the spontaneously broken symmetry in the Kekul\'e VBS phase is a discrete $C_3\!\cong\!\mathbb{Z}_3$ subgroup of the full $O(2)$ symmetry of the chiral XY GN(Y) Lagrangian, under which $\phi$ transforms as $\phi\!\rightarrow\!e^{2\pi ik/3}\phi$, $k\!=\!0,1,2$. As a result, a $\mathbb{Z}_3$ anisotropy term $\sim(\phi^3\!+\!\phi^{*3})$ is allowed in Eq.~(\ref{LGNY}), which may change the universality class of the transition or make it first order. In the $\epsilon$ expansion, such a term is strongly relevant at tree level, but quantum corrections at one-loop order render it irrelevant at the critical point in the large-$N_f$ limit~\cite{li2017}. Numerically, a continuous transition is found for $1\!\leq\!N_f\!\leq\!8$, and an emergent $O(2)$ rotational symmetry at criticality is observed for $N_f$ as low as 2~\cite{li2017}. An analogous transition is found in an $SU(4)$ Hubbard model on the $\pi$-flux square lattice, with columnar VBS order instead of a Kekul\'e pattern~\cite{zhou2018}, which doubles the unit cell in either the $x$ or $y$ directions. The broken symmetry in this case is the $C_4\!\cong\!\mathbb{Z}_4$ discrete rotation symmetry of the square lattice. A $\mathbb{Z}_4$ anisotropy term $\sim(\phi^4\!+\!\phi^{*4})$ is similarly allowed in the critical theory, but was shown to be irrelevant at criticality for all $N_f$ in the $\epsilon$ expansion~\cite{zerf2020}.

Other types of bond order are possible besides Kekul\'e VBS order. Mean-field theory predicts~\cite{raghu2008,weeks2010,grushin2013} that a large next-nearest-neighbor ($V_2$) interaction in the spinless Hamiltonian (\ref{HtV}) results in the formation of Haldane's quantum anomalous Hall (QAH) state~\cite{haldane1988}, with the spontaneous generation of imaginary next-nearest-neighbor hopping and ensuing $\mathcal{T}$-breaking. While one ED study~\cite{duric2014} has found signatures of the interaction-induced QAH state, the current consensus---supported by multiple ED~\cite{garcia-martinez2013,daghofer2014,capponi2015} and iDMRG~\cite{motruk2015} simulations---is that this state does not appear in the phase diagram of this model. For spin-1/2 fermions, mean-field theory predicts the extended Hubbard model (\ref{Hubbard}) with large $V_2$ can give rise to a quantum spin Hall (QSH) state~\cite{QSHE}, with spontaneously generated spin-orbit coupling~\cite{raghu2008}. Recent QMC studies have found interaction-induced semimetal-QSH quantum phase transitions using a designer Hamiltonian on the checkerboard lattice~\cite{he2018} and a fermionic Hamiltonian on the honeycomb lattice~\cite{liu2019}, in the $N_f\!=\!2$ chiral Ising and $N_f\!=\!1$ chiral Heisenberg universality classes, respectively.

\subsection{Deconfined criticality and symmetric mass generation}
\label{sec:DQCP}

In the Landau theory of phase transitions, direct transitions between phases with different broken symmetries are generically first order. An important class of quantum phase transitions are deconfined QCPs (DQCPs), such as the N\'eel-VBS transition of square-lattice quantum antiferromagnets, where a transition between distinct symmetry-breaking orders remains continuous, in violation of the Landau paradigm~\cite{senthil2004,senthil2004b}. While numerical investigations of DQCPs have traditionally focused on local spin Hamiltonians~\cite{sandvik2007}, the key ingredient for deconfined criticality---that the topological defects of one broken symmetry carry nontrivial quantum numbers under the other---naturally appears in Dirac fermion systems, when competing interactions favor orders described by mutually anticommuting mass terms~\cite{ghaemi2012}. Deep in either ordered phase, the intertwinement of the two broken symmetries is captured by a topological term in the nonlinear sigma model (NL$\sigma$M) obtained by integrating out the Dirac fermions~\cite{abanov2000,senthil2006}.

An early proposal~\cite{grover2008} for a DQCP with interacting Dirac fermions involves the competition between the spontaneous QSH order briefly discussed in Sec.~\ref{sec:Kekule}, which breaks the $SU(2)$ spin rotation symmetry, and $s$-wave spin-singlet superconductivity, which breaks the $U(1)$ particle-number conservation symmetry. For Dirac fermions, the three components of the QSH order parameter can be combined with the real and imaginary parts of the superconducting order parameter to form an $SO(5)$ vector $\b{\phi}$ that couples to five anticommuting mass matrices as in Eq.~(\ref{LGNY}). The resulting $SO(5)$ NL$\sigma$M contains a topological Wess-Zumino-Witten (WZW) term~\cite{witten1983} which suggests a possible QSH-superconductor DQCP. A Landau-forbidden continuous QSH-superconductor transition was indeed recently found in the QMC study of Ref.~\cite{liu2019}. Another example is a DQCP between the SDW and Kekul\'e VBS states discussed respectively in Sec.~\ref{sec:CDW_AF} and \ref{sec:Kekule}, which was found in QMC studies of designer~\cite{sato2017} and fermionic~\cite{li2019} Hamiltonians on the honeycomb lattice. Multicritical points at which the Dirac semimetal meets the two ordered phases were also recently studied in fRG and QMC~\cite{torres2019}.

We conclude this section by briefly mentioning an exotic class of transitions outside the GN(Y) paradigm of Sec.~\ref{sec:QPT}. They involve symmetric mass generation, whereby a many-body gap is opened up by sufficiently strong interactions in a Dirac fermion system, but in the absence of any spontaneous symmetry breaking. This scenario has antecedents in 1D~\cite{fidkowski2010}, and is realized in certain fermion lattice models with on-site interactions describing $N_f\!=\!4$ flavors of four-component Dirac fermions~\cite{slagle2015,ayyar2015,he2016}. Past a critical interaction strength, a weakly interacting Dirac semimetal phase transitions continuously to a gapped symmetric phase that, at strong coupling, is adiabatically connected to a product state of on-site many-body flavor singlets. A proposed critical theory for such transitions is a DQCP involving fractionalized fermion and scalar fields that interact with an emergent non-Abelian gauge field~\cite{you2018}.

\section{Dirac semimetal-superconductor transitions and emergent supersymmetry}
\label{sec:SUSY}

We have so far focused on transitions corresponding to particle-hole instabilities of a Dirac semimetal, in which the $U(1)$ particle-number conservation symmetry is preserved. We now focus on superconducting transitions, in which this symmetry is spontaneously broken to $\mathbb{Z}_2$. In this context, the spinor $\Psi$ appearing in Eqs.~(\ref{LGN},\ref{LGNY}) is a Nambu spinor, and the Dirac mass is the Bogoliubov quasiparticle gap.

The simplest superconductor of Dirac fermions, a conventional $s$-wave spin-singlet superconductor, can be realized in the attractive ($U\!<\!0$) nearest-neighbor Hubbard model (\ref{Hubbard}) on the half-filled honeycomb lattice~\cite{lee2009}, but is degenerate with a CDW state owing to an $SU(2)$ pseudospin symmetry~\cite{zhang1990}, as expected from particle-hole symmetry. Introducing a nonzero next-nearest-neighbor hopping $t'$ is expected to lift this degeneracy and stabilize the superconducting state. In mean-field theory, a transition from the Dirac semimetal to a gapped superconductor indeed obtains in this model at a critical interaction strength $|U|/t$~\cite{zhao2006,uchoa2007,kopnin2008}. The CDW can also be frustrated by considering the attractive Hubbard model on a non-bipartite lattice, such as the triangular lattice, which supports Dirac cones in the presence of alternating $\pi$ fluxes and likewise exhibits a superconducting transition~\cite{otsuka2018}.

The universality class of the quantum phase transition from Dirac semimetal to $s$-wave superconductor in the attractive Hubbard model is the chiral XY GN(Y) model with $N_f\!=\!2$ flavors of four-component Dirac fermions~\cite{roy2013}, already discussed in Sec.~\ref{sec:Kekule} in the context of Kekul\'e or columnar VBS semimetal-insulator transitions. In the latter context, the $U(1)$ symmetry at criticality was shown to be emergent, as the result of the irrelevance of $\mathbb{Z}_3$ or $\mathbb{Z}_4$ anisotropy terms, while in the former, it is the exact $U(1)$ particle-number conservation symmetry.

Another problem of particular recent interest has been the superconducting transition for an {\it odd} number of flavors of two-component Dirac fermions, as found on the surface of a 3D topological insulator. For a single flavor, the QCP is predicted to exhibit an emergent $\mathcal{N}\!=\!2$ supersymmetry (SUSY)~\cite{balents1998,lee2007,ponte2014,grover2014}. The corresponding chiral XY GNY Lagrangian (\ref{LGNY}) can be written explicitly as
\begin{align}\label{LSUSY}
\mathcal{L}=\overline{\psi}\slashed{\partial}\psi+|\partial_\mu\phi|^2+m^2|\phi|^2+\lambda^2|\phi|^4+h(\phi^*\psi_\uparrow\psi_\downarrow+\mathrm{h.c.}),
\end{align}
with $\psi\!=\!(\psi_\uparrow,\psi_\downarrow)$ the two-component Dirac spinor and $\phi$ the superconducting order parameter. In the $\epsilon$ expansion~\cite{balents1998,lee2007,ponte2014,grover2014,zerf2016,fei2016,zerf2017}, the fixed-point couplings at the GNY QCP obey $\lambda^2_*\!=\!h_*^2\!=\!\mathcal{O}(\epsilon)$. The Lagrangian (\ref{LSUSY}) is then invariant under $\mathcal{N}\!=\!2$ SUSY transformations, and is known as the (2+1)D $\mathcal{N}\!=\!2$ Wess-Zumino (WZ) model~\cite{aharony1997}. The Dirac fermion $\psi$ and Cooper pair field $\phi$ become superpartners, implying that their anomalous dimensions at criticality are equal to each other. SUSY is a highly constraining symmetry that allows a number of physical observables to be computed exactly at this QCP despite the presence of strong correlations. Owing to certain non-renormalization theorems in SUSY theories, the anomalous dimensions are known exactly: $\eta_\psi\!=\!\eta_\phi\!=\!1/3$~\cite{aharony1997}, which was verified in QMC simulations of this QCP~\cite{li2018}. Furthermore, the zero-temperature limit of the (real part of the) optical conductivity $\sigma(\omega,T)$, which is a universal constant at Lorentz-invariant QCPs~\cite{damle1997}, can be computed exactly in this case~\cite{witczak-krempa2016}:
\begin{align}
\sigma(\omega,0)=\frac{5(16\pi-9\sqrt{3})}{243\pi}\frac{e^2}{\hbar}\approx 0.2271\frac{e^2}{\hbar}.
\end{align}
Critical exponents not subject to non-renormalization theorems, such as the correlation length exponent $\nu$, have been computed in the $\epsilon$ expansion~\cite{balents1998,lee2007,ponte2014,grover2014,zerf2016,fei2016,zerf2017} and the conformal bootstrap~\cite{bobev2015}.

For a topological insulator surface with three flavors of two-component Dirac fermions, another type of SUSY was predicted to emerge at a quantum tricritical point between Dirac semimetal and nematic pair-density-wave (PDW) phases~\cite{jian2017b}. The fixed-point theory in this case is the so-called XYZ model~\cite{aharony1997}, unrelated to its quantum magnetism homonym and infrared dual to $\mathcal{N}\!=\!2$ SUSY QED$_3$. A QCP described by two copies of the $\mathcal{N}\!=\!2$ WZ model was also argued to arise at a Dirac semimetal-PDW transition in the spinless $V_1$-$V_2$ model on the honeycomb lattice described earlier (Sec.~\ref{sec:Kekule}), but with $V_1<0$~\cite{jian2015}.

Other studies have explored the possibility of realizing superconducting states with unconventional pairing symmetries in systems of interacting Dirac fermions. Spin-triplet and spin-singlet Kekul\'e PDW states were proposed as possible ground states of the Hubbard model on the honeycomb lattice with attractive on-site and nearest-neighbor interactions~\cite{roy2010}. Unconventional $d$-wave and extended $s$-wave pairings from repulsive interactions in the $\pi$-flux square lattice Hubbard model were also studied recently in QMC~\cite{guo2018}.

\section{Effect of disorder}
\label{sec:disorder}

Solid-state Dirac materials usually contain imperfections in the form of vacancies, impurities, or dislocations. Generally speaking, such quenched disorder may destabilize phases and phase transitions in otherwise clean samples, round first-order transitions into continuous ones, or produce new critical points. For sufficiently weak disorder, one may hope to capture its effect on a phase transition described by an effective low-energy Lagrangian by making the latter's coupling constants spatially random.

For an $O(n)$ symmetry-breaking quantum phase transition with order parameter field $\b{\phi}(\b{r},\tau)$ depending on spatial $\b{r}$ and imaginary time $\tau$ coordinates, the leading effects of disorder are captured by random-field or random-mass terms in the effective Lagrangian~\cite{Vojta2019}. In the first case, disorder breaks the $O(n)$ symmetry explicitly and introduces a linear term $-\b{h}(\b{r})\cdot\b{\phi}(\b{r},\tau)$ in the Lagrangian, where $\b{h}(\b{r})$ is a spatially random $O(n)$ vector field drawn from a particular distribution. Such disorder precludes spontaneous symmetry breaking in $d \leq 2$ spatial dimensions for discrete symmetries and in $d \leq 4$ for continuous symmetries~\cite{Imry1975,Aizenman1989,Greenblatt2009,Aizenman2012}. In the second case, disorder respects the $O(n)$ symmetry, and enters the effective Lagrangian as a spatially random correction to the scalar-field mass term, $\delta m^2(\b{r})\b{\phi}^2(\b{r},\tau)$. For short-range correlated disorder, $\langle\delta m^2(\b{r})\delta m^2(\b{r}')\rangle\!=\!W\delta(\b{r}-\b{r}')$,  the Harris criterion~\cite{Harris1974}, initially developed for classical phase transitions, answers the question whether the universality class of the transition changes with disorder. A clean QCP is stable against weak disorder if $\nu d\!>\!2$, where $\nu$ is the correlation length exponent in the clean limit.

While several studies have focused on disordered 2D Dirac fermions in the absence of interactions, or bosonic QCPs with disorder, the theoretical study of the effect of quenched disorder at Dirac QCPs is complicated by the interplay of Fermi statistics, spatial randomness, and strong correlations. The effect of disorder in the electron-electron attraction strength $g$ and chemical potential $\mu$ on the superconducting phase transition for $N\!=\!1$ flavor of two-component Dirac fermions, described in the clean limit by Eq.~(\ref{LSUSY}), was studied in Ref.~\cite{Nandkishore2013}. In the clean $g$-$\mu$ phase diagram at zero temperature, the semimetal phase is a segment $g \in(0,g_c)$ at zero chemical potential $\mu=0$ (see Sec.~\ref{sec:SUSY}). Weak disorder in the chemical potential creates puddles of electron- and hole-doped superconducting regions, which develop phase coherence below a critical temperature $T_c$ that, for long-range correlated disorder, is exponentially weak in the disorder strength. For short-range correlated disorder, mean-field theory predicts that $T_c$ is doubly exponentially weak in the disorder strength. However, the inclusion of rare-region effects brings back the single exponential dependence of $T_c$ on the disorder strength. Thus, the semimetal phase is destabilized by chemical potential disorder. These conclusions are consistent with a later numerical study of the attractive Hubbard model (\ref{Hubbard}) on a honeycomb lattice subjected to a random scalar potential~\cite{Potirniche2014}. For magnetic disorder, which preserves $U(1)$ particle-number conservation symmetry but breaks $\mathcal{T}$, a mean-field theory with a treatment of disorder in the self-consistent Born approximation predicts a universal rate of logarithmic increase of $g_c$ with dimensionless disorder strength $\gamma$, $d\ln g_c/d\gamma\!=\!3\ln 2$~\cite{Ozfidan2016}, which can be interpreted as a quantum-critical analog of the Abrikosov-Gor'kov universal $T_c$ suppression rate $dT_c/d\Gamma\!=\!-\pi/4$~\cite{abrikosov1961}, with $\Gamma$ the magnetic scattering rate.

In Ref.~\cite{Nandkishore2013}, a perturbative $\epsilon$-expansion study in $4\!-\!\epsilon$ spacetime dimensions was also applied to reveal the effect of short-range correlated random-mass disorder on the $N\!=\!1$ Dirac semimetal-superconductor QCP at $\mu\!=\!0$, $g\!=\!g_c$. Averaging over disorder via the replica trick~\cite{QPT} adds an effective interaction nonlocal in time,
\begin{align}\label{Sdis}
S_\text{dis}=-\frac{W}{2} \sum_{ab}\int d^d\b{r}\,d \tau\,d \tau'\,
|\phi_a|^2(\b{r},\tau)|\phi_b|^2(\b{r},\tau'),
\end{align}
to the action $S=\int d^d\b{r}\,d\tau\,\mathcal{L}$, where $\mathcal{L}$ is the $N_f\!=\!1/2$ chiral XY GNY Lagrangian (\ref{LSUSY}) and $a,b$ are replica indices. In this approach, the disorder is relevant at the clean QCP---in agreement with the Harris criterion, since there $\nu\!\approx\!0.917\!<\!1$~\cite{bobev2015}---but subsequently exhibits a runaway flow to strong coupling, placing the ultimate fate of the QCP beyond the reach of perturbative RG. Technically, this happens because the engineering scaling dimensions of the $|\phi|^4$ and Yukawa couplings in Eq.~(\ref{LSUSY}) are $\mathcal{O}(\epsilon)$, while that of the disorder-induced interaction (\ref{Sdis}) is $\mathcal{O}(1)$. To remedy this problem, in Ref.~\cite{Yerzhakov2018} a double epsilon expansion~\cite{Dorogovtsev1980, Boyanovsky1982, Lawrie1984} was applied to the chiral XY GNY model with $N_f$ four-component Dirac fermion flavors in the presence of short-range correlated random-mass disorder. In this expansion, perturbative RG calculations are performed with dimensional regularization in $4\!-\!\epsilon$ spatial and $\epsilon_\tau$ imaginary time dimensions, which makes all relevant couplings have $\mathcal{O}(\epsilon,\epsilon_\tau)$ scaling dimensions at tree level. At one-loop order, it was shown in this approach that the clean GNY fixed point is stable against disorder for $N_f\!\geq\!1$, with the disorder being marginally relevant for $N_f\!=\!1/2$. These findings are in accord with the Harris criterion. However, for $N_f\!>\!\frac{1}{2}$ it was found that beyond a critical disorder strength $W_c\!\sim\!\mathcal{O}(\epsilon,\epsilon_\tau)$, the superconducting transition is governed by a new finite-randomness fermionic QCP, with critical couplings $\lambda_*^2$, $h_*^2$, and $W_*\!>\!W_c$ all nonzero. This is not in contradiction with the Harris criterion, which characterizes the stability of an existing clean QCP against weak disorder but does not preclude the formation of new disordered QCPs at strong disorder. For $N_f\!=\!2$, corresponding to spinful fermions on the honeycomb lattice, the disordered QCP at $W_*$ and the multicritical point at $W_c$ merge into a single, marginally stable fixed point. All these finite-randomness fermionic QCPs are characterized by non-Gaussian critical exponents, with nonzero anomalous dimensions $\eta_\phi$ and $\eta_\psi$. Furthermore, the dynamic critical exponent is found to obey $z\!\neq\!1$, as expected since disorder breaks the Lorentz invariance of the clean Lagrangian (\ref{LSUSY}). (For an example of a Dirac QCP with $z\!\neq\!1$ even in the clean limit, see Ref.~\cite{christou2020}.)

Another feature of these disordered fermionic QCPs is the possibility of unusual RG flows. For $N_f \geq 7/2$ the system of beta functions linearized near the disordered fixed points possesses complex-conjugate eigenvalues. Such a phenomenon was previously found for certain critical points of classical disordered magnets~\cite{Dorogovtsev1980, Boyanovsky1982, Lawrie1984,Khmelnitskii1978} as well as a (clean) superconducting QCP in 3D Luttinger semimetals~\cite{boettcher2016}. It leads to a spiraling flow towards a stable-focus fixed point, which manifests itself in predicted oscillatory corrections to scaling in various physical properties~\cite{Khmelnitskii1978}. Questions such as whether these flows occur in reality or are merely an artefact of the double epsilon expansion~\cite{Goldman2019}, and whether rare-region effects produce additional quantum Griffiths phases in the vicinity of the QCP~\cite{Vojta2019}, can only be answered by non-perturbative approaches, e.g., numerical studies.

Besides the effect of random-mass disorder on the Dirac semimetal-superconductor transition, the effects of random hopping and random gauge fields on Dirac semimetal-insulator quantum phase transitions were studied using RG methods in Ref.~\cite{Foster2006,vafek2008}. We also mention recent studies~\cite{Goswami2017,Thomson2017,zhao2017,Lee2019} of the effect of disorder in QED$_3$, of potential relevance to quantum phase transitions out of the Dirac spin liquid, the subject of the following section.

\section{Transitions out of the Dirac spin liquid}
\label{sec:DSL}

As mentioned at the end of Sec.~\ref{sec:DiracCMP}, the Dirac spin liquid (DSL) is an exotic ground state of frustrated 2D quantum antiferromagnets where spin-1 magnons fractionalize into spin-1/2 fermionic spinons with a massless Dirac dispersion. By contrast with transitions out of the Dirac semimetal considered so far, quantum critical phenomena in the DSL are affected not only by the gapless fermionic nature of this phase but also by its fractionalized character. The emergent gauge fields in the DSL enter as additional soft modes at a putative QCP, and produce new classes of critical behavior distinct from the pure GN(Y) universality classes discussed in Sec.~\ref{sec:SM-I}-\ref{sec:SUSY}.

\subsection{U(1)-{N\'eel} transition}\label{sec:U1Neel}

A sign-problem-free QMC study of a designer Hamiltonian for the DSL on the 2D square lattice provided a concrete realization of this phase~\cite{Xu2019}. The Hamiltonian considered describes spin-1/2 fermions on the sites of this lattice interacting with $U(1)$ rotors on nearest-neighbor bonds,
\begin{align}
H  = \frac{J}{4}\sum_{\langle ij\rangle}L_{ij}^{2}-t\sum_{\langle ij\rangle\sigma}\left(c_{i\sigma}^{\dagger}e^{i\theta_{ij}}c_{j\sigma}+\text{h.c.}\right)
 +K\sum_{\Box}\cos\left(\boldsymbol{\Delta}\times\boldsymbol{\theta}\right).\label{eq:Hamiltonian}
\end{align}
Here, $c_{i\sigma}^{\left(\dagger\right)}$ annihilates (creates) at site $i$ a fermion with spin $\sigma\!=\!\uparrow,\downarrow$, to be understood as a fermionic spinon. Likewise, $\theta_{ij}\!\in\!\left[0,2\pi\right)$ and $L_{ij}$, a pair of canonically conjugate rotor operators on bond $ij$, behave as a lattice $U(1)$ gauge potential and a lattice electric field, respectively. The coefficient $K\!>\!0$ of the magnetic-field term produces a background magnetic flux $\boldsymbol{\Delta}\!\times\!\boldsymbol{\theta}\!=\!\pi$ per plaquette ($\Box$), as in Fig.~\ref{fig:1}(c), thus generating $N_f\!=\!2$ flavors of massless four-component Dirac fermions at low energies (see Sec.~\ref{sec:DiracCMP}). The coefficient $J$ of the electric-field term controls the strength of $U(1)$ gauge fluctuations above this background, which minimally couple to the Dirac ``spinons''.

At small $J$, a power-law correlated $U(1)$ deconfined phase is found, which is adiabatically connected to the DSL state of frustrated spin-1/2 quantum antiferromagnets~\cite{kim1999,rantner2001,rantner2002,hermele2004}. The low-energy theory of the DSL is QED$_3$ with $N_f\!=\!2$ flavors of massless four-component Dirac fermions. Noncompact QED$_3$ is known to flow to a conformally invariant fixed point for sufficiently large $N_f$~\cite{appelquist1988,nash1989}. On a lattice, the gauge field is compact, which allows for monopoles (instantons) and possible confinement by monopole proliferation~\cite{polyakov1975,PolyakovBook,polyakov1977}. Provided $N_f$ is sufficiently large, monopole proliferation is absent~\cite{Pufu2014}, but the threshold value of $N_f$ for this effect is not known precisely. The numerical results of Ref.~\cite{Xu2019} suggest this threshold value may be as low as 2, but finite-size effects prevent the establishment of a rigorous bound.

Besides the DSL phase itself, upon increasing $J$ a continuous transition to a N\'eel state is found at a critical value $J\!=\!J_c$, in analogy to the antiferromagnetic SDW transition for the spin-1/2 Hubbard model on the $\pi$-flux square lattice (Sec.~\ref{sec:CDW_AF}). The existence of this transition is expected because, as for the latter model in the large-$U$ limit, Eq.~(\ref{eq:Hamiltonian}) reduces in the large-$J$ limit to the spin-1/2 antiferromagnetic Heisenberg model, which orders antiferromagnetically at zero temperature on the square lattice. However, due to the fractionalized nature of the paramagnetic phase, the critical theory is not the chiral Heisenberg GNY model discussed in Sec.~\ref{sec:CDW_AF}, but rather a gauged version thereof, the chiral Heisenberg QED$_3$-GNY model~\cite{ghaemi2006}:
\begin{align}\label{eq:QEDGN}
\mathcal{L}_\text{DSL-N\'eel}=\overline{\Psi}\slashed{\partial}\Psi+\mathcal{L}_\b{\phi}+\frac{1}{4}F_{\mu\nu}^2
+h\boldsymbol{\phi}\cdot\overline{\Psi}\b{\sigma}\Psi+ie\overline{\Psi}\slashed{A}\Psi,
\end{align}
where $\Psi=(\Psi_\uparrow,\Psi_\downarrow)$ denotes an $SU(2)$ doublet of four-component Dirac fermions and $\b{\sigma}$ is a vector of spin Pauli matrices. As for the SDW transition of Dirac semimetals, the order parameter is a real $O(3)$ scalar field $\b{\phi}$ governed by the $O(3)$ Lagrangian $\mathcal{L}_\b{\phi}$, but in addition there is a $U(1)$ gauge field $A_\mu$ governed by the Maxwell Lagrangian and coupled minimally to the fermions with a nonzero gauge coupling $e$.

The critical properties of Eq.~(\ref{eq:QEDGN}) were studied in Ref.~\cite{Dupuis2019} at one-loop order in the $\epsilon$ expansion, and in Ref.~\cite{Zerf2019} at four-loop order in the $\epsilon$ expansion as well as in the large-$N_f$ expansion. Those studies established there is a stable QCP with exponents differing from those
of both conformal QED$_{3}$ and the pure chiral Heisenberg GNY QCP, owing to the presence of two kinds of soft bosonic modes at criticality---critical antiferromagnetic spin waves and gauge fluctuations.

\subsection{U(1)--valence-bond-solid transition}\label{sec:U1VBS}

In Refs.~\cite{Xu2019,wang2019,janssen2020}, the Hamiltonian (\ref{eq:Hamiltonian}) was extended to $N_f\!>\!2$ flavors of fermions, $\sigma\!\rightarrow\!\alpha=1,\ldots,N_f$, and studied in QMC. For $N_f\!=\!4,6,8$, this extended Hamiltonian again exhibits a continuous quantum phase transition from a gapless $U(1)$ deconfined phase at small $J$, described by conformal QED$_3$ with $N_f$ flavors of four-component Dirac fermions, to an ordered phase at $J\!>\!J_c$. However, this time the ordered phase is found to be a quantum paramagnet with columnar VBS order. As for the DSL-N\'eel transition just discussed, the critical theory of this transition, derived in Ref.~\cite{Boyack2019b}, is a gauged analog of the critical theory for the appropriate semimetal-insulator transition---here the columnar VBS transition for the $SU(N_f)$ Hubbard model on the $\pi$-flux square lattice~\cite{zhou2018}, already discussed in Sec.~\ref{sec:Kekule}. The resulting theory is the chiral XY QED$_3$-GNY model, obtained by adding a Maxwell term for the emergent $U(1)$ gauge field to the chiral XY GNY Lagrangian (\ref{LGNY}), and covariantizing the derivative $\slashed{\partial}\!\rightarrow\!\slashed{\partial}\!+\!ie\slashed{A}$ in the fermion kinetic term. Since the chiral XY GNY model admits two equivalent formulations (see Sec.~\ref{sec:Kekule}), so are there two equivalent formulations of the critical theory for the $U(1)$-VBS transition~\cite{Boyack2019b}, with the choice $\b{\mathcal{M}}\!=\!(i\Gamma_3,i\Gamma_5)$ usually referred to as the chiral XY or $O(2)$ QED$_3$-GNY model~\cite{Xu2019}, and $\b{\mathcal{M}}\!=\!(1,i\gamma_5)$ corresponding to the gauged NJL model~\cite{klevansky1989}. A $\mathbb{Z}_4$ anisotropy term is allowed in the critical theory, as for the semimetal-VBS transition on the $\pi$-flux square lattice, but was shown to be irrelevant at the QCP~\cite{zerf2020}.

The critical properties of the chiral XY QED$_3$-GNY model have been studied in the $\epsilon$ expansion at one-loop~\cite{scherer2016} and four-loop~\cite{zerf2020} orders, and in the large-$N_f$ expansion at leading~\cite{Gracey1993a,Boyack2019b} and subleading~\cite{zerf2020} orders.

\subsection{Gapped spin liquids and deconfined criticality}

The transitions discussed in Secs.~\ref{sec:U1Neel} and \ref{sec:U1VBS} are confinement transitions, whereby gapping out the Dirac spinons triggers monopole proliferation below the fermion gap. The N\'eel and VBS phases obtained for $J\!>\!J_c$ are indeed conventional phases without fractionalization, which should not support propagating emergent gauge bosons. However, it is in principle possible to open a spinon gap in such a way that the resulting pure gauge theory at low energies remain in a deconfined phase; one then obtains a topologically ordered spin liquid, which is gapped but still fractionalized. A putative quantum phase transition from a DSL to a gapped $\mathbb{Z}_2$ spin liquid~\cite{wen1991} driven by spinon pairing, which Higgses the $U(1)$ gauge field, was studied using $\epsilon$ and large-$N_f$ expansions in Ref.~\cite{boyack2018}. This transition can be seen as a gauged version of the Dirac semimetal-superconductor transition discussed in Sec.~\ref{sec:SUSY}. The corresponding critical theory is a type of chiral XY QED$_3$-GNY model distinct from that discussed in Sec.~\ref{sec:U1VBS}, and would be more appropriately termed the Higgs-QED$_3$-GNY model. By contrast with the critical theory of the $U(1)$-VBS transition, here the complex order parameter $\phi$ carries nontrivial gauge charge because it corresponds to a Cooper pair amplitude of spinons; thus critical properties differ in the two theories.

Another transition of this type is between a DSL and a chiral spin liquid~\cite{kalmeyer1987}, driven by the spontaneous generation of a $\mathcal{T}$-breaking QAH mass for the Dirac spinons~\cite{wen1989} in a gauged analog of Haldane mass generation in Dirac semimetals (Sec.~\ref{sec:Kekule}). A candidate critical theory for this transition---the chiral Ising QED$_3$-GNY model with $N_f\!=\!2$ flavors of four-component Dirac fermions, a gauged version of the chiral Ising GNY model discussed in Sec.~\ref{sec:CDW_AF}---has been studied in both the $\epsilon$~\cite{janssen2017,Ihrig2018b,Zerf2018} and large-$N_f$~\cite{gracey1992,gracey2018b} expansions. However, the $\mathcal{T}$-breaking QAH mass is an $SU(2N_f)$ singlet operator~\cite{kubota2001}, while the Lagrangian considered in those studies only has an $SU(N_f)\!\times\!SU(N_f)$ symmetry~\cite{Boyack2019,Benvenuti2019}. The correct $SU(2N_f)$-symmetric Lagrangian was studied in the large-$N_f$ expansion in Ref.~\cite{Boyack2019}, and its critical properties were shown to differ from those of the Lagrangian with $SU(N_f)\!\times\!SU(N_f)$ symmetry due to the presence of Aslamazov-Larkin diagrams, previously considered only in pure QED$_3$~\cite{hermele2007,chester2016}.

For $N_f\!=\!1$, or equivalently $N\!=\!2$ flavors of two-component Dirac fermions, the critical $SU(2)$-symmetric chiral Ising QED$_3$-GNY model was recently conjectured to be infrared dual to the N\'eel-VBS DQCP of square-lattice quantum antiferromagnets~\cite{Wang2017}. As one predicted consequence of this duality, the anomalous dimensions of the N\'eel and VBS order parameters at the bosonic DQCP should match the scalar-field anomalous dimension $\eta_\phi$ on the fermionic side of the duality, and the scaling dimension of the flavor-adjoint fermion bilinear in the fermionic theory should be identical with $3\!-\!\nu^{-1}$, where $\nu$ is the correlation length exponent of the bosonic DQCP. Pad\'e and Pad\'e-Borel resummations of the large-$N_f$ exponents for the chiral Ising QED$_3$-GNY model in Ref.~\cite{Boyack2019} are reasonably consistent with those equivalences, using critical exponents at the N\'eel-VBS DQCP obtained from large-scale QMC simulations~\cite{sandvik2007,Melko2008,Nahum2015,Nahum2015b}. However, whether the N\'eel-VBS transition is in fact a (D)QCP or a weakly first-order transition is still a matter of active debate~\cite{ma2019,nahum2019,sandvik2020}.

\section{Conclusion}
\label{sec:conclusion}

The emerging field of Dirac quantum criticality is situated at the confluence of two major themes in modern condensed matter physics---Dirac matter~\cite{wehling2014,vafek2014} and strong correlations. Besides having profound connections to various other areas of condensed matter physics, such as topological materials and frustrated magnetism, Dirac quantum criticality also deeply relates to many topics of current research in high-energy physics, including 3D conformal field theories, supersymmetry, and field-theoretic dualities. On the experimental front, recent progress in identifying correlated materials with nontrivial band topology~\cite{witczak-krempa2014,schaffer2016,rachel2018} and realizing systems of ultracold fermions on optical honeycomb lattices with tunable interactions~\cite{uehlinger2013,greif2015} may lead in the near future to the observation of the exotic phenomena discussed here.

\begin{acknowledgement}
We acknowledge support from NSERC (grant \#RGPIN-2014-4608), the CRC Program, CIFAR, Alberta Innovates, the University of Alberta, and its Theoretical Physics Institute.
\end{acknowledgement}

\section*{Author contribution statement}

All authors contributed significantly to the writing of the manuscript. Sections 6-7 largely review research carried out by H.Y. and R.B. and supervised by J.M.

\bibliography{diracQPT}
\bibliographystyle{epj}

\end{document}